\documentclass[11pt,twoside]{article}
\usepackage{asp2004}
\usepackage{epsf}
\usepackage{graphics}
\usepackage{lscape}
\usepackage{natbib}

\def\Mpc{\,{\rm Mpc}}
\def\HO{{100h\,{\rm km\,sec^{-1}\,Mpc^{-1}}}}

\def\fun#1#2{\lower3.6pt\vbox{\baselineskip0pt\lineskip.9pt
  \ialign{$\mathsurround=0pt#1\hfil##\hfil$\crcr#2\crcr\sim\crcr}}}

\input epsf

\def\plotone#1{\centering \leavevmode
\epsfxsize= 0.6\columnwidth \epsfbox{#1}}

\def\be{\begin{equation}}
\def\ee{\end{equation}}
\def\ba{\begin{eqnarray}}
\def\ea{\end{eqnarray}}
\def\bea{\begin{eqnarray}}
\def\eea{\end{eqnarray}}

\def\edcomment#1{\iffalse\marginpar{\raggedright\sl#1\/}\else\relax\fi}
\reversemarginpar

\begin{document}
\bibliographystyle{apj}
\title{Weak Lensing and Supernovae:  Complementary Probes of Dark
Energy}
\author{L. Knox, A. Albrecht \& Y.S. Song}
\affil{University of California, Davis Department of Physics, One
Shields Avenue, Davis, California 95616, USA}

\begin{abstract}
Weak lensing observations and supernova observations, combined
with CMB observations, can both provide powerful constraints on
dark energy properties.  Considering statistical errors only, we
find luminosity distances inferred from 2000 supernovae and
large-scale ($l < 1000$) angular power spectra inferred from
redshift-binned cosmic shear maps place complementary constraints
on $w_0$ and $w_a$ where $w(z) = w_0 + w_a(a-1)$. Further, each
set of observations can constrain higher-dimensional
parameterizations of $w(z)$; we consider eigenmodes of the $w(z)$
error covariance matrix and find such datasets can each constrain
the amplitude of about 5 $w(z)$ eigenmodes.  We also consider
another parameterization of the dark energy.
\end{abstract}
\thispagestyle{plain}

\section{Introduction}

In the past half-decade a great variety of methods have emerged
for `observing dark energy', enough to fill an entire 4-day
meeting.  One might ask why we need such a great variety of
methods.  To this question we give three answers:
\begin{itemize} \item Independent methods provide the ultimate
systematic error test.  \item Independent methods provide for less
model-dependent probes of cosmic acceleration.
\item Different methods constrain different directions in the parameter space.
\end{itemize}

Given the importance of the dark energy mystery and the challenges
to constraining its properties, this diversity of methods is a
blessing.   In this talk we concentrate on two methods: weak
lensing and supernovae.  Both of these highly different types of
observations are potentially powerful probes of dark energy. Here
we examine item 3: the complementary nature of the statistical
errors.

Weak lensing observations can deliver enormously rich datasets; there
are a multitude of ways of using these data to constrain dark
energy. Here we concentrate solely on what can be done with the data
on large scales because this is simplest to model, most robust to
increases in shape noise and spurious psf power, and because others at
this meeting (Bernstein and Jain) discussed the potentially powerful
uses of smaller-scale data.  See, for example,
\cite{refregier03,takada03b,tyson02,jain03,bernstein04,song03,hu03a}.

In Section 2 we present our models of the datasets and in Section
3 the constraints on $w_0$ and $w_a$. In Section 4 we examine
constraints on higher-dimensional parameterizations of $w(z)$ as
well as discuss another parameterization of the dark energy.  In
Section 5 we conclude.

\section{Models of the Data}

In this talk we assume a CMB survey, a cosmic shear survey and a
supernova survey are used to constrain a large parameter space.
Although we are only interested in the dark energy parameters
here, the cosmic shear survey in particular is sensitive to many
other parameters and so we must simultaneously fit for them as
well.  The CMB survey is very useful for constraining these
non-dark-energy parameters.  We assume Planck (comprehensively 
described in \citep{planck} and, more specifically for our calculations, in 
\citet{song03}) as our CMB survey since we expect these data to be
significantly more constraining than WMAP data \citep{bennett03} (due mostly to
Planck's higher angular resolution) and these data will be
available by the time we have any cosmic shear and supernova
datasets like those we describe below.

As in \citet{kaplinghat03b} and \citet{song03} 
we take our (non-$w(z)$) set to be
${\cal P} = \{\omega_m, \omega_b, \omega_\nu, \theta_s, z_{\rm
ri},k^3P_\Phi^i(k_f),n_s,n_s',y_{\rm He}\}$, with the assumption
of a flat universe. The first three of these are the densities
today (in units of $1.88\times 10^{-29}{\rm g}/{\rm cm}^3$) of
cold dark matter plus baryons, baryons and massive neutrinos. We
assume two massless species and one massive species. The next is
the angular size subtended by the sound horizon on the
last--scattering surface. The Thompson scattering optical depth
for CMB photons, $\tau$, is parameterized by the redshift of
reionization $z_{\rm ri}$. The primordial potential power spectrum
is assumed to be $k^3P_\Phi^i(k) = k_f^3P_\Phi^i(k_f)(k/k_f)^{n_s
-1+n_S'ln(k/k_f)}$ with $k_f = 0.05\Mpc^{-1}$. The fraction of
baryonic mass in Helium (which affects the number density of
electrons) is $y_{\rm He}$.  We Taylor expand about ${\cal
P}=\{0.146,0.021,0,0.6,6.3,6.4\times 10^{-11},1,0,0.24\}$. The
Hubble constant for this model is $h=0.655$ where $H_0 = \HO$.

\subsection{Cosmic Shear Data Model}

We call our fiducial weak lensing survey G$2\pi$ because we
imagine it as a ground-based survey of half of the sky. We assume
a galaxy redshift distribution for a limiting magnitude in R of 26
inferred from observations with the Subaru telescope
\cite{nagashima02}. The shape of this distribution is
well-described by the following analytic form: \bea
dn/dz &\propto& z^{1.3}\exp\left[-\left(z/1.2\right)^{1.2}\right] \ \ \ {\rm for } \ z <1 \nonumber \\
dn/dz &\propto & z^{1.1}\exp\left[-\left(z/1.2\right)^{1.2}\right]
{\rm for } \ z >1.\eea We use this distribution with the
modification that half of the galaxies in the $1.2 < z < 2.5$
range are discarded as undetectable.
The amplitude of the distribution is such that, after
this cut, the number density of galaxies is 65 per sq.
arcmin\footnote{J.A. Tyson, private communication}.

We further assume that the galaxies can be divided, by photometric
redshift estimation, into eight different redshift bins: [0-0.4],
..., [2.8-3.2] and that for $40 < l < 1000$ systematic errors are
negligible.  Note that this last assumption is a very strong one
and much work will be necessary to make it a valid one.  Finally,
we assume that the shape noise (expressed as a per-component rms
shear) is given by $\gamma_{\rm rms}(z) = 0.15+0.035z$.

For more details of the data modeling, see \citet{song03}.

\begin{figure}[!ht]
\label{fig:shear-shear} \plotone{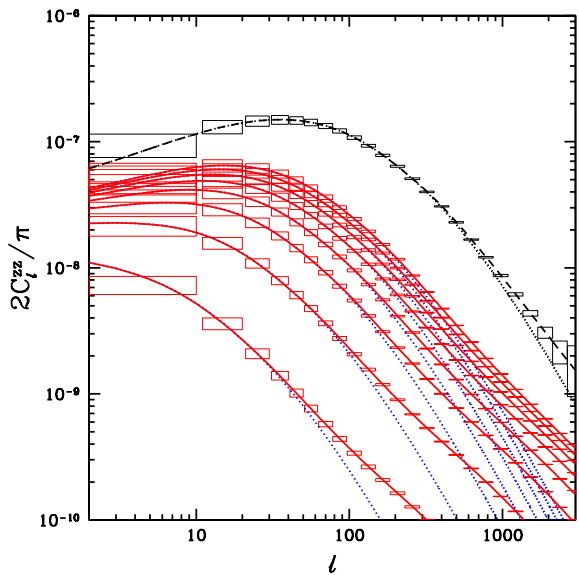} \caption{The
shear-shear auto power spectra. The 8 solid curves are the shear
power spectra from each of the galaxy source planes, $B_1$ to
$B_8$.  Dotted curves are the linear perturbation theory
approximation.  From bottom to top the source plane redshift
ranges are $B_1$: $z\in[0.0,0.4]$, $B_2$: $z\in[0.4,0.8]$, $B_3$:
$z\in[0.8,1.2]$, $B_4$: $z\in[1.2,1.6]$, $B_5$: $z\in[1.6,2.0]$,
$B_6$: $z\in[2.0,2.4]$, $B_7$: $z\in[2.4,2.8]$ and $B_8$:
$z\in[2.8,3.2]$. The error boxes are forecasts for G4$\pi$ (see
Table I). The top dashed curve is the shear power spectrum for the
CMB source plane.  The error boxes are forecasts for CMBpol.}
\end{figure}

The eight shear auto power spectra can be determined with high
accuracy over a large range in $l$, as is shown in
Figure~\ref{fig:shear-shear}.  In addition to the auto power
spectra shown, there are also 9(9-1)/2=36 cross spectra that one
can measure.  We include the cross spectra in our parameter error
forecasts.  They do not add much statistical weight because their
statistical errors are highly correlated with the errors in the
auto spectra.  The large number of largely redundant 2-point
functions will be useful for revealing the contaminating influence
of systematic errors.  See, for example, \citet{takada03c}.

\subsection{Supernova Data Model}

For the supernova survey we assume 2000 distributed in redshift as
described in \citet{kim04} as a baseline SNAP supernova survey.  In
addition, we assume measurement of 100 local supernovae.  To our
parameter set, detailed above, we add a supernova luminosity
calibration parameter.  We do not explicitly include systematic
errors in our analysis. The supernova forecasts in some sense do
include systematic error estimates, because the SNAP baseline
survey, by design, does not attempt to reduce statistical errors
far below the expected systematic error limits (by intentionally
restricting the number of supernovae in each redshift interval).

\section{Constraints on $w_0$ and $w_a$}

\begin{figure}[!ht]
\label{fig:wcon} \plottwo{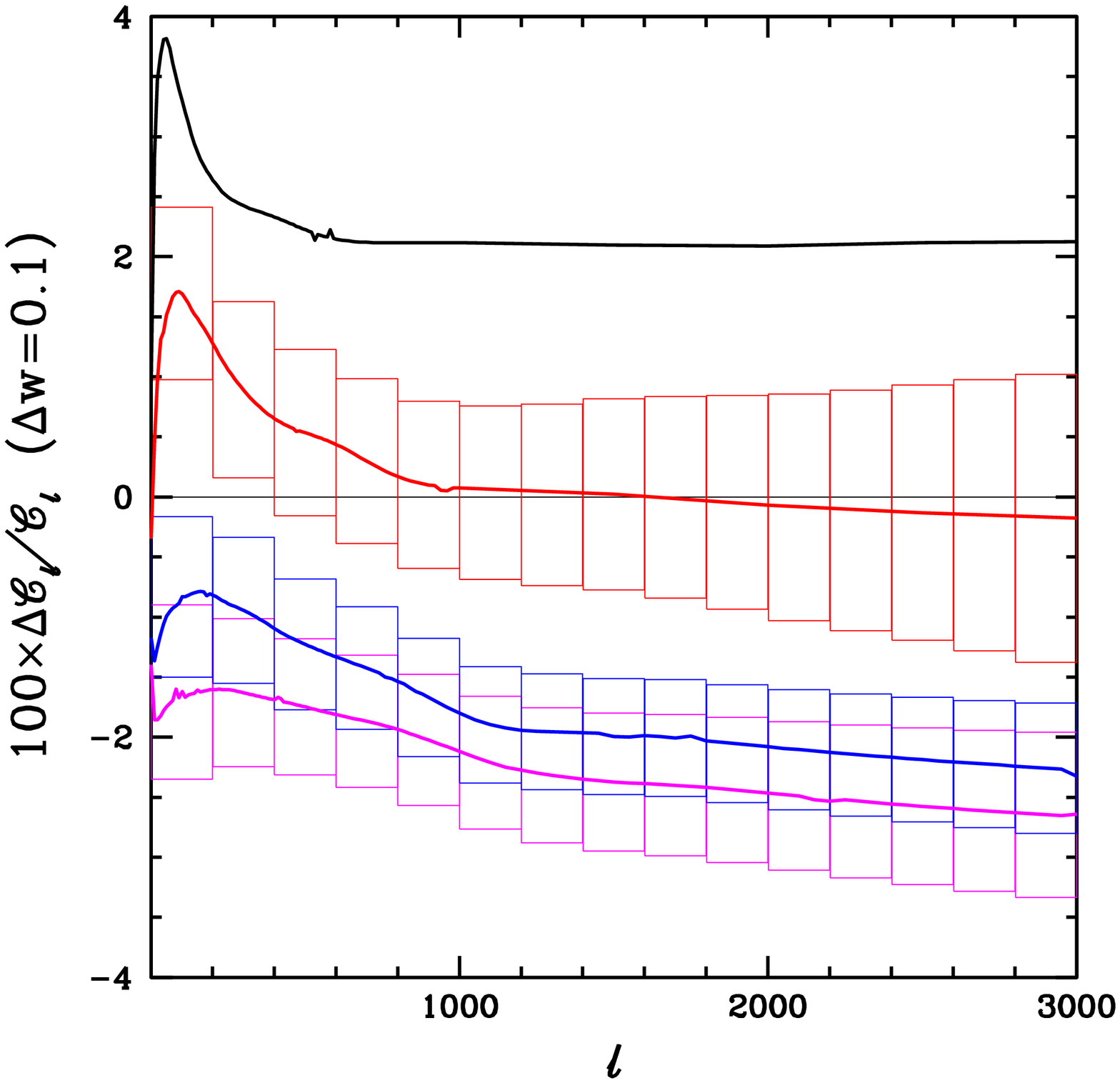}{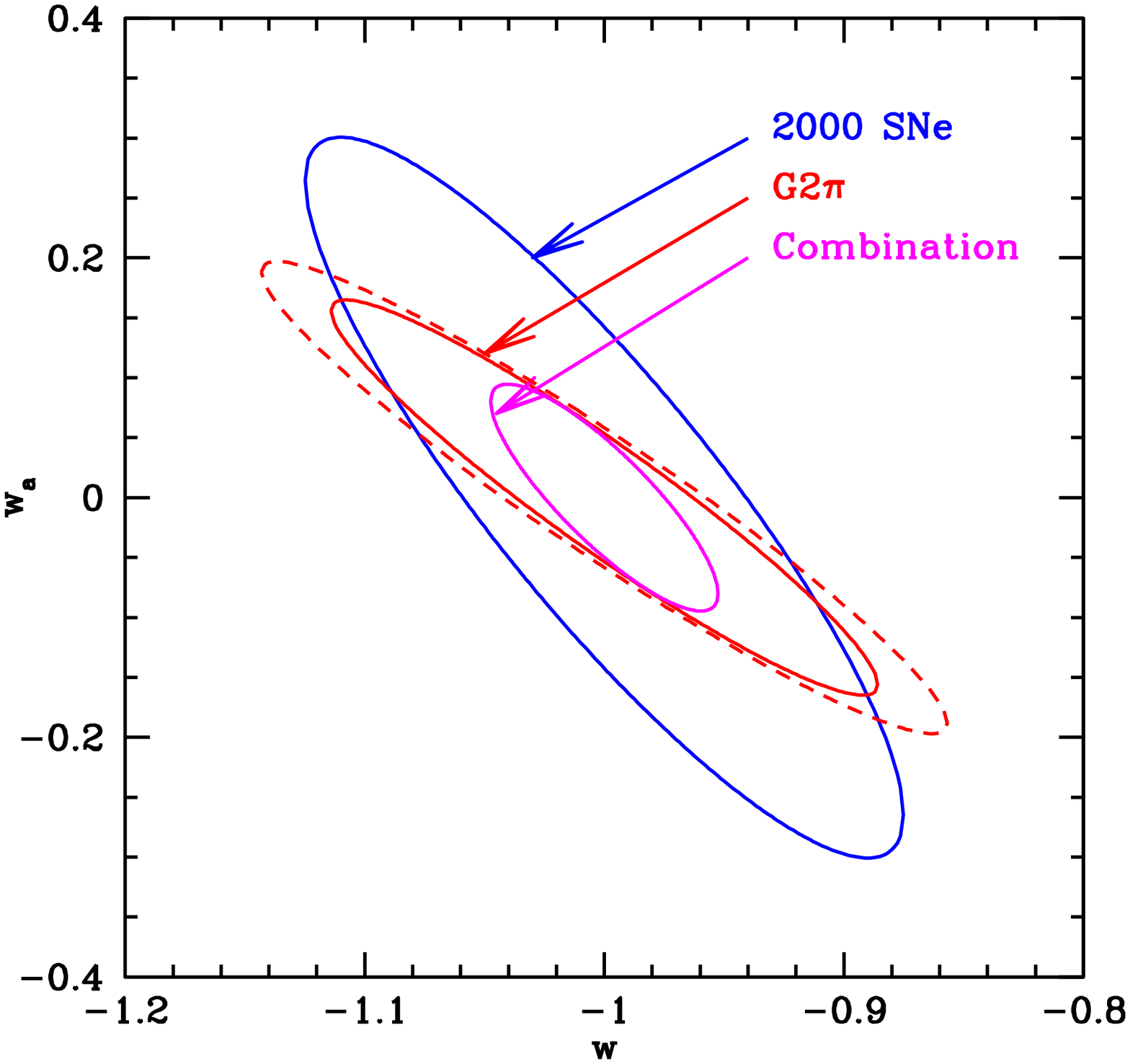} \caption{Left
panel shows difference between $w=-.9$ and $w=-1$ shear power
spectra for source bins centered at $z=0.2$, $0.6$, $1.4$ and
$3.0$ from top to bottom.  The $z=0.2$ curve has no error bars on
it because their large extent would clutter the graph. The right
panel shows one $\sigma$ error contours in the $w_0$-$w_a$ plane
for G$2\pi$, 2000 SNe and the combination (as labeled). The
dashed curve is for G$2\pi$ with the source density uniformly
decreased by a factor of 2.}
\end{figure}

For dark energy models with sound speeds near unity, the Jeans
scale is on the horizon and we can ignore clustering of the dark
energy on scales smaller than the horizon, and certainly at $l <
40$, the lowest $l$ value we consider.  Thus the dark energy
results in a time-dependent but scale-independent suppression of
power.  However, the scale-independence in three dimensions does
not translate into a scale-independence of the two-dimensional
projection to shear power spectra because the projection folds the
time and scale-dependence together.  Note though that in the
special case of a power-law power spectrum the projection does
preserve the scale-dependence. Thus, in the left panel of
Figure~\ref{fig:wcon} we see changing $w$ changes the shape of the
power spectra, but not at small scales where the three-dimensional
power spectrum is well--approximated as a power law.  In addition
to altering the growth rate, dark energy alters the shear power
spectra by changing $D_A(z)$ and therefore the projection of the
fluctuation power at different redshifts onto the sky.

The trend with redshift in the left panel of Figure~\ref{fig:wcon}
can be understood as arising from the dependence of the growth
rate on $w_0$.  Recall that we are changing $w_0$ while keeping
the angular size of the sound horizon on the last-scattering
surface (and hence the angular-diameter distance to
last-scattering) fixed. Increasing $w_0$ from $w_0=-1$ means more
dark energy at high redshift (since the dark energy density is now
increasing with redshift as opposed to being constant) and less
dark energy at low redshift (in order to keep the angular-diameter
distance fixed).  Thus, growth is suppressed at high redshift and
enhanced at low redshift.

The scale-dependence of the shear power spectra dependence on the dark
energy is a boon.  The scale-dependence prevents the dark energy
effects from being completely degenerate with a redshift-dependent
calibration error. We have ignored shear calibration uncertainty in
our analysis.  It would be interesting to include it (as in
\citet{ishak04}) and see how effectively the dataset can simultaneously
constrain calibration and cosmological parameters, just with the 
assumption that the calibration is scale-independent.

\newpage

The supernova data constrain the dark energy through the
dependence of the luminosity distance (equivalent to $D_A(z)$ in a
flat Universe) on the dark energy. The resulting constraint from
the 2000 supernovae are also shown in Fig.~\ref{fig:wcon}.  We see
that the two error ellipses are not perfectly aligned.  Combining
the datasets results in a factor of 2 decrease in the $w_a$
direction and a factor of 3 decrease in the $w_0$ direction.

An advantage of using only the large angular scale statistics is
their robustness to increases in shape noise. At $l < 1000$, for
most of the source redshift bins, the G$2\pi$ shear errors are
dominated by sample variance not shape noise variance. Thus an
increase in the shape noise variance by a factor of 2 for all
source bins (as would happen if the source density were uniformly
lower than expected by a factor of two) only leads to a small
increase in the error contours, as shown with the dashed contours
in Figure~\ref{fig:wcon}.

\section{Constraints on higher-dimensional parameterizations}

\begin{figure}[!ht]
\label{fig:eig}
\plotfiddle{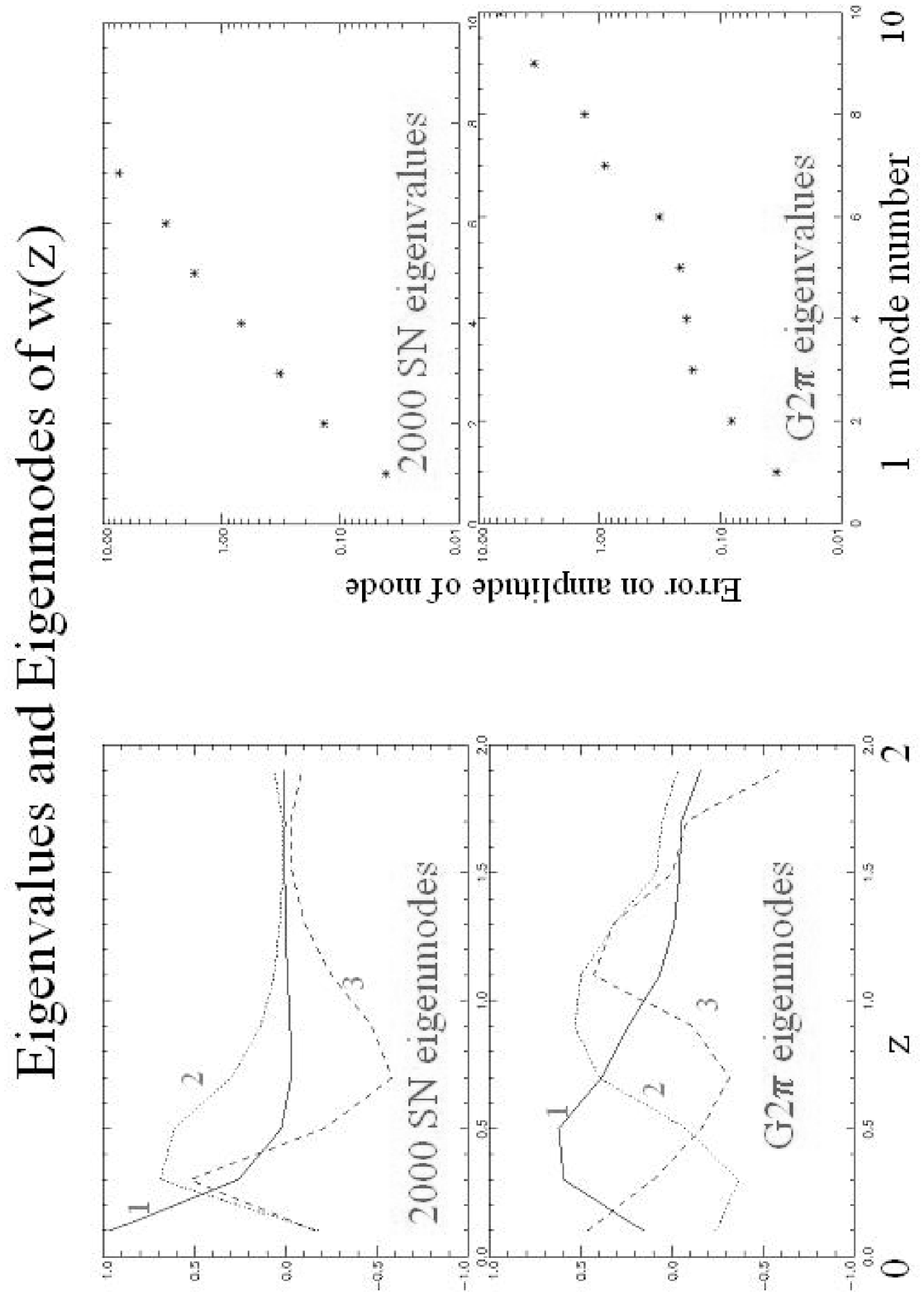}{4.5in}{-90}{65}{65}{-260}{370}
\caption{Eigenvalues (right) and first three eigenmodes (left) of
the $w(z)$ error covariance matrix for G$2\pi$+Planck and 2000
SNe+Planck. The large contributions to the eigenmodes from the
highest redshift-bin are an artifact of that bin being much
broader than the rest, extending all the way to the
last-scattering surface.}
\end{figure}

To deepen our understanding of how these surveys are
constraining dark energy, we have examined how they constrain the
function $w(z)$, rather than its simple parameterization by $w_0$
and $w_a$. We proceed by binning $w(z)$ in redshift bins and then
identifying the eigenmodes and eigenvalues of the binned $w(z)$
error covariance matrix as was done for supernovae by \citet{huterer03}.

In the left panels of Figure~\ref{fig:eig} we plot the three
eigenmodes with the best determined eigenvalues.  We see a
striking difference in the modes for 2000 SNe vs. the modes for
G$2\pi$:  those for G$2\pi$ stretch out to higher redshifts. The
reason for this is that lensing is less sensitive to the growth
factor at the lower redshifts where the source density in a given
redshift bin is small and the lensing window (for sources at
higher redshift) is also small.  Thus the supernovae are better at
detecting changes in $w(z)$ at lower redshift and G$2\pi$ tends to
be better at detecting changes at higher redshift.

G$2\pi$ and 2000 SNe also have strikingly different eigenvalue
spectra.  The error on the amplitude of the best determined mode
is quite similar for each ($\sim 0.03$).  But the 2000 SNe
spectrum is much steeper.  G$2\pi$ has seven modes with errors
smaller than unity, whereas 2000 SNe has four.

To further explore the power of these experiments to constrain
dark energy we chose eigenmodes of $w(z)$ because it was
calculationally convenient for us.  There are probably
parameterizations with superior qualities.  One we have begun
exploring is modes of $\rho_{d.e.}(z)$, with the constraint that
one of the modes is constant; i.e., independent of $z$.  This approach has
two virtues.  First, the density is more directly related to the
data than the equation-of-state parameter.  Second, interpretation
of data as evidence for non-cosmological-constant behavior is
straightforward;  one only need assess how significantly non-zero
the amplitudes of the non-constant modes are.

\section{Conclusions}

As we have seen from Bernstein's talk and Jain's talk, weak lensing
datasets are very rich, with dark energy constraints possible from a
variety of statistical measures. Here we have studied just one
statistic, the shear-shear two-point functions on large angular
scales.  Even restricting ourselves in this manner, we find the
potential power of large-area weak lensing surveys is extraordinary.
Further, we see that the statistical weight of the data constrain
directions in parameter space different from those constrained by
supernovae; i.e., they are complementary.  Therefore the combination
is particularly powerful.

To more fully understand how these different observations probe
dark energy we examined eigenmodes of the $w(z)$ error covariance
matrix.  We saw that the G$2\pi$ observations are better at
probing $w(z)$ at high redshift and the supernova observations are
better at lower redshifts.  A handful of eigenmodes could be
determined with less than 100\% errors.

This work has illustrated the different statistical
uncertainties on dark energy parameters for two types of
probes.  A side by side comparison of comprehensive error
forecasts is a much more ambitious undertaking that would require
detailed consideration of systematic errors and the extent to
which they can be controlled.

\acknowledgments We thank B. Gold, M. Kaplinghat, E. Linder,
S. Perlmutter, U. Seljak and T. Tyson for useful conversations and the
organizers for a stimulating meeting.  This material is based upon
work supported by the DoE, NASA grant NAG5-11098 and by NSF grant
0307961.

\bibliography{/work3/knox/bib/cmb3}

\begin{thebibliography}{}
%\expandafter\ifx\csname natexlab\endcsname\relax\def\natexlab#1{#1}\fi

\bibitem[{{Bennett} {et~al.}(2003){Bennett}, {Halpern}, {Hinshaw}, {Jarosik},
  {Kogut}, {Limon}, {Meyer}, {Page}, {Spergel}, {Tucker}, {Wollack}, {Wright},
  {Barnes}, {Greason}, {Hill}, {Komatsu}, {Nolta}, {Odegard}, {Peiris},
  {Verde}, \& {Weiland}}]{bennett03}
{Bennett}, C.~L., {Halpern}, M., {Hinshaw}, G., {Jarosik}, N., {Kogut}, A.,
  {Limon}, M., {Meyer}, S.~S., {Page}, L., {Spergel}, D.~N., {Tucker}, G.~S.,
  {Wollack}, E., {Wright}, E.~L., {Barnes}, C., {Greason}, M.~R., {Hill},
  R.~S., {Komatsu}, E., {Nolta}, M.~R., {Odegard}, N., {Peiris}, H.~V.,
  {Verde}, L., \& {Weiland}, J.~L. 2003, \apjs, 148, 1

\bibitem[{{Bernstein} \& {Jain}(2004)}]{bernstein04}
{Bernstein}, G. \& {Jain}, B. 2004, \apj, 600, 17

\bibitem[{{Hu} \& {Jain}(2003)}]{hu03a}
{Hu}, W. \& {Jain}, B. 2003, ArXiv Astrophysics e-prints

\bibitem[{{Huterer} \& {Starkman}(2003)}]{huterer03}
{Huterer}, D. \& {Starkman}, G. 2003, Physical Review Letters, 90, 31301

\bibitem[{{Ishak} {et~al.}(2004){Ishak}, {Hirata}, {McDonald}, \&
  {Seljak}}]{ishak04}
{Ishak}, M., {Hirata}, C.~M., {McDonald}, P., \& {Seljak}, U. 2004, \prd, 69,
  083514

\bibitem[{{Jain} \& {Taylor}(2003)}]{jain03}
{Jain}, B. \& {Taylor}, A. 2003, Physical Review Letters, 91, 141302

\bibitem[{{Kaplinghat} {et~al.}(2003){Kaplinghat}, {Knox}, \&
  {Song}}]{kaplinghat03b}
{Kaplinghat}, M., {Knox}, L., \& {Song}, Y. 2003, ArXiv Astrophysics e-prints,
  3344

\bibitem[{{Kim} {et~al.}(2004){Kim}, {Linder}, {Miquel}, \& {Mostek}}]{kim04}
{Kim}, A.~G., {Linder}, E.~V., {Miquel}, R., \& {Mostek}, N. 2004, \mnras, 347,
  909

\bibitem[{{Nagashima} {et~al.}(2002){Nagashima}, {Yoshii}, {Totani}, \&
  {Gouda}}]{nagashima02}
{Nagashima}, M., {Yoshii}, Y., {Totani}, T., \& {Gouda}, N. 2002, ArXiv
  Astrophysics e-prints, 7483

\bibitem[{{Refregier} {et~al.}(2003){Refregier}, {Massey}, {Rhodes}, {Ellis},
  {Albert}, {Bacon}, {Bernstein}, {McKay}, \& {Perlmutter}}]{refregier03}
{Refregier}, A., {Massey}, R., {Rhodes}, J., {Ellis}, R., {Albert}, J.,
  {Bacon}, D., {Bernstein}, G., {McKay}, T., \& {Perlmutter}, S. 2003, ArXiv
  Astrophysics e-prints

\bibitem[{{Song} \& {Knox}(2003)}]{song03}
{Song}, Y. \& {Knox}, L. 2003, ArXiv Astrophysics e-prints

\bibitem[{{Takada} \& {Jain}(2003)}]{takada03b}
{Takada}, M. \& {Jain}, B. 2003, ArXiv Astrophysics e-prints

\bibitem[{{Takada} \& {White}(2003)}]{takada03c}
{Takada}, M. \& {White}, M. 2003, ArXiv Astrophysics e-prints

\bibitem[{{Tauber}(2001)}]{planck}
{Tauber}, J.~A. 2001, in IAU Symposium, 493--+

\bibitem[{Tyson {et~al.}(2003)Tyson, Wittman, Hennawi, \& Spergel}]{tyson02}
Tyson, J.~A., Wittman, D.~M., Hennawi, J.~F., \& Spergel, D.~N. 2003,
  astro-ph/0209632. Nuc. Phys. B, 124, 21

\end{thebibliography}

\end{document}